\documentclass[10pt,conference]{IEEEtran}

\pdfoutput=1
\usepackage{etex}

\usepackage[utf8]{inputenc}
\usepackage[english]{babel}
\usepackage[bookmarks=false,hidelinks]{hyperref}
\usepackage{balance}
\usepackage{tabularx}
\usepackage{xcolor}
\usepackage{pgfplots}
\usepackage[autolanguage]{numprint}
\usepackage{xspace}
\usepackage{graphicx}
\usepackage{multirow}
\usepackage{booktabs}
\usepackage{stfloats}
\usepackage{pifont}
\usepackage[inline]{enumitem}

\pgfplotsset{compat=1.14}

\usepackage{algorithm}
\usepackage{algorithmic}

\makeatletter
\newcommand{\ALC@uniqueautorefname}{Line}
\makeatother
\AtBeginDocument{%
}

\definecolor{bblue}{HTML}{4F81BD}
\definecolor{rred}{HTML}{C0504D}
\definecolor{ggreen}{HTML}{9BBB59}
\definecolor{ppurple}{HTML}{9F4C7C}

\newcommand{\mytilde}{\raise.17ex\hbox{$\scriptstyle\mathtt{\sim}$}}

\newcommand{\answer}[2]{\vspace{.3cm}{\centering\fbox{\parbox{0.95\columnwidth}{\textbf{Answer to RQ#1}. #2}}}\vspace{.3cm}}

\usepackage{ifthen}

\newboolean{showcomments}
\setboolean{showcomments}{true}

\ifthenelse{\boolean{showcomments}}
  {\newcommand{\nb}[3]{
  {\color{#2}\small\fbox{\bfseries\sffamily\scriptsize#1}}
  {\color{#2}\sffamily\small$\triangleright~$\textit{\small #3}$~\triangleleft$}
  \GenericWarning{}{LaTeX Warning: #1: #3}
  }
  }
  {\newcommand{\nb}[3]{}
  }

\newcommand{\bugswarm}[0]{BugSwarm\xspace}
\newcommand{\aprbugs}[0]{112\xspace}
\newcommand{\nbPairs}[0]{\numprint{2949}\xspace}

\usepackage{amssymb,amsmath}

\author{
    \IEEEauthorblockN{
      Thomas Durieux,
      Rui Abreu
    }
\IEEEauthorblockA{INESC-ID and IST, University of Lisbon, Portugal}
}

\title{Critical Review of \bugswarm for Fault Localization and Program Repair}

\begin{document}

\maketitle

\begin{abstract}
Benchmarks play an essential role in evaluating the efficiency and effectiveness
of solutions to automate several phases of the software development lifecycle. 
Moreover, if well designed, they also serve us well as an
important artifact to compare different approaches amongst themselves. \bugswarm
is a benchmark that has been recently published, which contains \numprint{3091}
pairs of failing and passing continuous integration builds. According to the authors, the benchmark has
been designed with the automatic program repair and fault localization
communities in mind. Given that a benchmark targeting these communities ought to
have several characteristics (e.g., a buggy statement needs to be present), we
have dissected \bugswarm to fully understand whether the benchmark suits
these communities well. Our critical analysis shows several limitations in
the benchmark: only \aprbugs/\numprint{3091} (\numprint{3.6}\%) are suitable to
evaluate techniques for automatic fault localization or program repair.
\end{abstract}

\section{Introduction}
Empirical software engineering focuses on gathering evidence, mainly via
measurements and experiments involving software artifacts. The information
collected is then used to form the basis of theories about the processes under
study. Benchmarks, therefore,  play an essential role in the empirical studies,
as, in many situations, they are the only source of information to evaluate new
approaches. In particular, benchmarks are used to evaluate approaches to testing 
(e.g., automatic test generation), automatic fault localization
(FL), and automatic program repair (APR).  These two research fields
require benchmark of (behavioral) bugs to evaluate the precision of the fault
localization and the ability to generate correct patches, respectively.

In the past year, the research community put a lot of effort to develop new
benchmarks for those fields. Indeed, four new benchmarks have been presented in
the past months: BUGSJS~\cite{gyimesi2019bugsjs},
Bears~\cite{Madeiral2019Bears}, Code4Bench~\cite{majd2019code4bench} and
\bugswarm~\cite{dmeiri2019bugswarm}. They complement the existing benchmarks:
Defects4J~\cite{Just2014Defects4J},
IntroClass~\cite{LeGoues2015ManyBugsIntroClass},
ManyBugs~\cite{LeGoues2015ManyBugsIntroClass},
IntroClassJava~\cite{Durieux2016IntroClassJava},
Bugs.jar~\cite{Saha2018BugsDotjar}, QuixBugs~\cite{Lin2017QuixBugs}. As the
number of benchmarks grows, it is becoming more and more important to have a
clear picture of the characteristics of each benchmark. This is important to
ensure the quality of the evaluations that use such benchmarks and to guide
researchers to the benchmarks that suit the best their needs.

In this paper, we present a critical review of the \bugswarm benchmark, recently
published at the International Conference on Software Engineering (ICSE'19)~\cite{dmeiri2019bugswarm}. 
\bugswarm is a benchmark of \numprint{3091} pairs of
failing and passing builds, designed for the automatic fault localization and 
program repair communities. They succeed to reproduce \numprint{3.05}\%
(\numprint{3091}/\numprint{101265}) of the pairs of builds they considered. They ended up
with \numprint{1827} pairs of Java builds and \numprint{1264} pairs of Python
builds that are between one and four years old. Those builds are extracted from
108 Java projects and 52 Python projects, with an average of \numprint{19.3}
pairs of builds per project.

In this paper, we characterize the human patches, the failures and the usage
of \bugswarm. We then focus our analysis on the applicability of \bugswarm in
evaluating state of the art automatic fault localization and program repair 
research.

In our analysis, we identify that an important number of pairs of builds
are ill suited for automatic fault localization and program repair fields.
Indeed, we observe that \bugswarm contains, e.g., duplicate commits and builds 
that fail due to non-behavioral problems. Our analysis shows that only
\aprbugs/\numprint{3091} (a mere \numprint{3.6}\%) meet the criteria of those
research fields. Reporting the suitable builds per programming language shows 50
entries for Java and 62 for Python only.

The difference of number is explained by the fact that \bugswarm is a benchmark
of failing and passing builds and not a benchmark of bugs. Builds can fail for
multiple others reasons than regression bugs. Automatic program repair and fault
localization rely on bugs for their evaluation. Using all the builds
in \bugswarm would result in bad repair and fault localization precision rate and 
misleading analysis.

The contributions of this paper are:
\begin{itemize}
  \item a critical review of the \bugswarm benchmark with respect to automatic 
  fault localization and program repair;

  \item a characterization of \bugswarm's content in terms of build execution, diff analysis, failure type and patch type;

  \item a set of lessons learned for designing future benchmarks for automatic program repair and fault localization;

  \item the source code of the \numprint{3091} \bugswarm pairs of builds, the
  TravisCI's logs for each build and the diffs between the passing and failing
  builds;

  \item an interactive website~\cite{website} to browse and filter \bugswarm's pairs of builds.
\end{itemize}

This paper is an analysis of the content and usage of \bugswarm. This analysis
provides guidelines for researchers on how to use \bugswarm and prevent misusage
or incorrect recommendations.

The remainder of this paper is organized as follows. \autoref{sec:background}
presents the background of this paper: the \bugswarm benchmark and the requirements of automatic program repair and fault localization tools. \autoref{sec:contributions} contains our analysis
of the \bugswarm benchmark. \autoref{sec:discussion} discusses the lessons learned of this analysis
and the threats to validity. \autoref{sec:rw} presents the related works and \autoref{sec:conclusion} concludes this paper.

\section{Background}\label{sec:background}

\subsection{What is \bugswarm?}\label{sec:bugswarm}

\begin{figure*}[t]
  \centering
  \includegraphics[width=.9\textwidth]{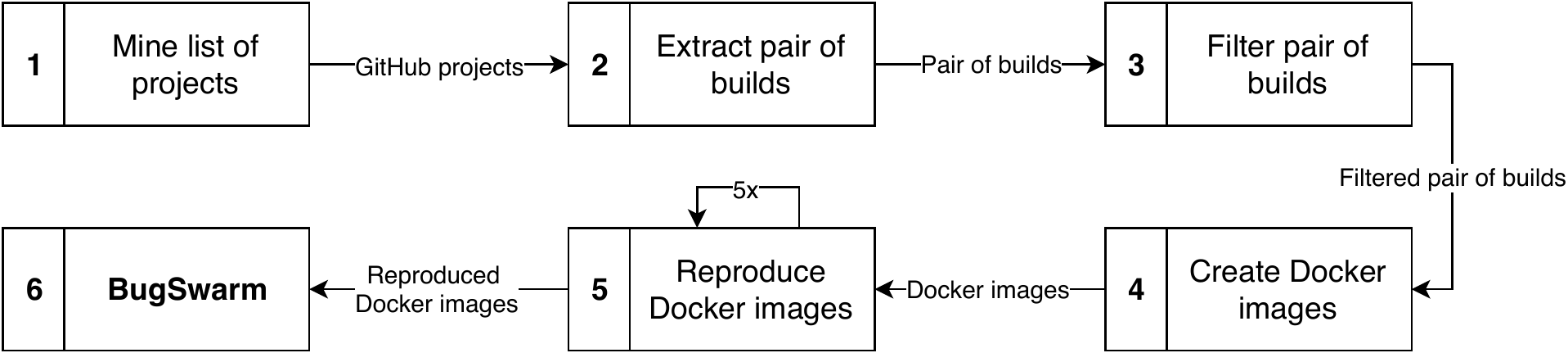}
  \caption{Methodology followed to create the \bugswarm benchmark.}
  \label{fig:BugSwarm_architecture}
\end{figure*}

\bugswarm is a benchmark of \numprint{3091} pairs of builds that have been
published by Dmeiri et al. at ICSE'19~\cite{dmeiri2019bugswarm}.

The idea of \bugswarm is to collect pairs of failing/passing builds from a
continuous integration service (in this case TravisCI). The failing build
contains the incorrect behavior to fix and the passing build contains the oracle
of the fix, e.i. the human modification that makes the failing build pass. This
approach is also used by Bears benchmark~\cite{Madeiral2019Bears} but differs
from the benchmarks that are traditionally used by automatic program repair and
fault localization fields such as Defects4J~\cite{Just2014Defects4J} and
QuixBugs~\cite{Lin2017QuixBugs}.

Each pair of builds is composed of a failing build and a passing builds that are
encapsulated in a Docker image. The Docker image provides the source code and
the scripts to reproduce the passing and failing execution. \bugswarm is
accompanied with a command line interface that downloads and starts the Docker
images as well as an infrastructure to execute scripts inside the Docker image.

\autoref{fig:BugSwarm_architecture} presents the six main steps that \bugswarm's
authors followed to create it.
\begin{enumerate}
  \item \textit{Mine a list of GitHub projects}. The first step consists of
  collecting a set of GitHub projects that used TravisCI.

  \item \textit{Extract pair of builds}. During the following step, they analyze the 
  build history of each project and collect all pairs of failing/passing
  builds. The complexity of this task resides in the ability to recreate the
  correct history of the branches since Git history is a tree and TravisCI
  history is linear. They needed an additional step to match the git history with
  the TravisCI history.

  \item \textit{Filter pair of builds}. The next step consists of keeping only
  the pairs of builds that have a chance to be reproducible. In this case, they
  only consider the builds where TravisCI uses a Docker image to run the build.

  \item \textit{Create Docker images}. The fourth step is to create a new Docker
  image that contains the two builds (failing and passing) and the scripts that
  execute them.

  \item \textit{Reproduce Docker images}. This fifth step is an important step.
  It consists of executing five times each Docker image that has been created at
  the previous step to ensure that the behaviors of the builds are consistent.
  They parsed the TravisCI logs and the logs produced by the new Docker image to
  extract the number of failing tests. They consider that the builds are
  reproduced when the number of failing test is identical.

  \item \textit{\bugswarm}. The final step is to create an infrastructure that
  makes it possible to use the \bugswarm and its \numprint{3091} pairs of
  builds.
\end{enumerate}

\autoref{tab:bugswarm_metrics} presents the main metrics that have been
presented in \bugswarm's paper. \bugswarm contains \numprint{3091} pairs of
builds, \numprint{1827} from Java applications, \numprint{1264} from Python
applications. Those builds are coming from \numprint{160} different GitHub
projects, 108 Java projects, 52 Python projects. The usage information is
available on \bugswarm's website: \url{http://bugswarm.org} and for further
details, we recommend reading the \bugswarm paper~\cite{dmeiri2019bugswarm}.

\begin{table}[t]
  \caption{The original metrics presented in \bugswarm's paper.}
  \label{tab:bugswarm_metrics}
  \centering
  \begin{tabularx}{\linewidth}{@{}X r r r@{}}
    \toprule
                       & Java    & Python & All \\\midrule
    \# Pairs of Builds & \numprint{1827} & \numprint{1264} & \numprint{3091} \\
    \# Projects       & 108 & 52 & 160    \\
    \bottomrule
  \end{tabularx}
\end{table}

\subsection{APR and FL Requirements}\label{sec:requirements}

The current state of the art of automatic program repair and fault localization techniques have a set of requirements for the buggy programs that they receive as inputs. 
We identify two categories of requirements.
The first category contains the requirements related to the execution itself and the second category is to ensure the fairness of empirical evaluations.

The requirements for automatic program repair and fault localization:
\begin{enumerate}
  \item \textbf{Bug type.} Current APR and FL techniques only target behavioral bugs that are present in the source code of the program.
  It means that bugs in configuration or external files are currently not compatible with APR and FL.
  
  \item \textbf{Specification of the program.} The test-suite of the application is currently being used as the main specification in APR and FL. The passing tests specify the correct behavior of the application and a failing test describes the incorrect behavior.

  \item \textbf{Program setup.} The execution setup of the program has to be known such as the path of the sources, path of the tests, path of the binary, the classpath (for Java) and version of the source.
  
  \item \textbf{Uniqueness of the bug.} The requirement is important to ensure that the technique does not overfit one specific bug and consequently introduce bias in the analysis of the results.
  
  \item \textbf{Human patch.} The human patch is currently being used in APR and FL evaluation as the perfect oracle.
  It provides the solution on how to fix the bug and provide the location of the bug.
\end{enumerate}

Those requirements define the necessary conditions to be able to use and evaluate state-of-the-art 
APR and FL approaches on bugs.

\section{\bugswarm Analysis}\label{sec:contributions}
This section contains our analysis of \bugswarm.

\subsection{Research Questions}

In this study, we address the following three research questions:
\begin{itemize}
    \item[\textbf{RQ1}.] What are the main characteristics of \bugswarm's pairs
    of builds regarding the requirements for FL and APR? In this research question, we analyze the pairs of builds using
    three different axes: 1. failing builds, 2. the human patches, and 3. the
    failure reasons in order to identify pairs of builds that match APR and FL requirements.
    \item[\textbf{RQ2}.] What is the execution and storage cost of \bugswarm? In the second
    research question, we describe and analyze the usage of \bugswarm and we
    estimate the execution and storage cost of running an experiment with \bugswarm on Amazon cloud.
    \item[\textbf{RQ3}.] Which pairs of builds meet the requirements of
    Automatic Program Repair (APR) and Fault Localization (FL) techniques? In
    the final research question, we put in perspective \bugswarm's pairs of
    builds with the state of the art automatic program repair and fault
    localization techniques and we identify which ones could be used by those
    fields.
\end{itemize}

\subsection{Protocol}\label{sec:protocol}

In this section, we present the protocol that we followed to collect the artifacts used to answer our three research questions. 
We identify the following four artifacts that need to be collected:
\begin{enumerate}
  \item The source code of each pair of builds.
  \item The diff between the failing and passing build.
  \item The build information from TravisCI (including the execution logs).
  \item The Docker image information from DockerHub.
\end{enumerate}

\begin{figure}
  \includegraphics[width=0.47\textwidth]{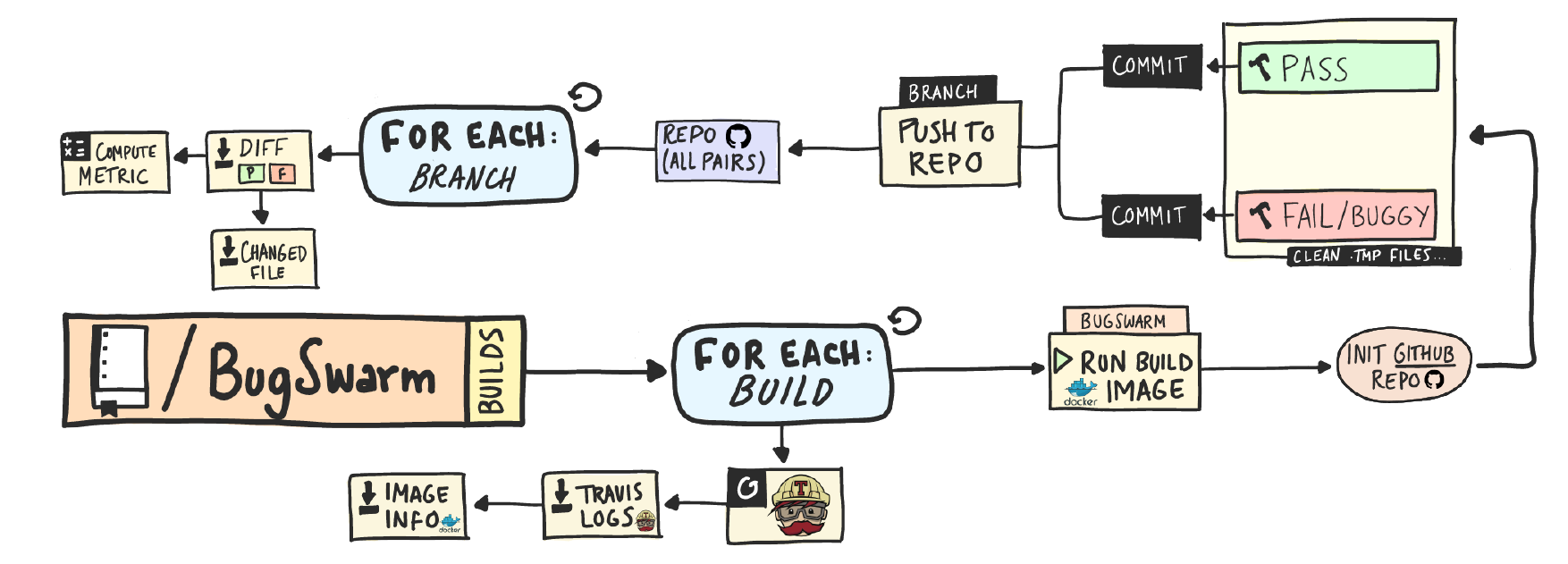}
  \caption{The protocol that we follow to extract the buggy and passing builds from \bugswarm's Docker images and the protocol to extract the metrics that we used to answer the research questions.}
  \label{fig:protocol}
\end{figure}

\autoref{fig:protocol} describes our protocol.
We first get all the pairs of builds information using \bugswarm's API.\footnote{\bugswarm's API request to get the list of pairs of builds: \url{http://www.api.bugswarm.org/v1/artifacts/?where={"reproduce_successes":{"$gt":4,"lang":{"$in":["Java","Python"]}}}}}
and we iterate over it. For each build, we use
the \bugswarm command line to run the Docker image. Inside the Docker image, we first prepare
the buggy and passed version of the application by removing the temporary files that had been introduced during the creation of \bugswarm. The temporary files are duplicate versions of the repository.
Then, we create a new Git repository where we commit the buggy and passed
version. The repository is then pushed to our GitHub repository in its
branch \cite{repo}. Then, we download the failing and passing log execution from
TravisCI and the Docker manifest of the image via DockerHub.

Once all the pairs of builds are pushed on GitHub, we download the diffs between
each failing/passing version using GitHub's API.
Finally, we compute the metrics on the collected artifacts such as the number of
changed files, the number of changed lines, the types of file that have been
changed or the time between the failing and passing commits. All the collected
artifacts, the scripts to collect and analyze are publicly available on our
GitHub repository~\cite{repo}. The repository also offers the access of the source of each pair of 
build; this is a rather convenient when the execution of the builds is not require for the analysis. 
We also created a website that presents the human diffs and the collected metrics for each artifact~\cite{website}.

\subsection{RQ1. Characteristics of \bugswarm's Pairs of Builds}\label{sec:rq1}

In this section, we analyze \bugswarm from three different angles:
\begin{enumerate*}
  \item a general analysis of the pairs of builds,
  \item an analysis of the human patch and its diff,
  \item and an analysis of the failures.
\end{enumerate*}
The analyses used the data that is collected
with the protocol described in \autoref{sec:protocol}.

\subsubsection{Buggy Builds}

For the first angle, we do a general analysis of the pairs of builds of  \bugswarm.
\autoref{tab:buggy_metric} presents the main metrics of this analysis.
This table is divided into four columns. 
The first column presents the name of the metric, the second column contains the number for Java pairs of builds, the third for Python and the last one for Java and Python pairs of builds.

Our first observation is that the number of builds reported in this paper (\nbPairs) and \bugswarm's paper (\numprint{3091}) are different (line one and two of \autoref{tab:buggy_metric}).
Indeed, we considered all pairs of builds that are reproduced successfully five times like it is described in \bugswarm's paper (see Section 4-B in \cite{dmeiri2019bugswarm}).
Surprisingly, \bugswarm authors did not consider their criteria in their final selection of the pairs of builds and consequently the reported number is in contradiction with the paper. 
We observe that the number changes with the time between \numprint{2042} and \numprint{2949}. It is possible that the API will not respond the same number in the future.
We contacted to authors about this point and they told us that the property had been overwritten by mistake in the database and they fixed it manually.

Our second observation is that \numprint{40.08}\% ((\nbPairs $-$ \numprint{1767})/\nbPairs) of the builds have a duplicate failing commit id. 
It means that those \numprint{40.08}\% should not be considered by the approaches that only consider the source code of the application otherwise it introduces misleading results (see requirement 4 in \autoref{sec:requirements}). 
It also shows that Java pairs of builds are slightly more impacted by the duplicate commits compared to Python builds (\numprint{41.26}\% vs. \numprint{38.48}\%). 
Finally, we observed that eight Docker images are unavailable: 
Adobe-Consulting-Services-acs-aem-commons-315891915,
Adobe-Consulting-Services-acs-aem-commons-358605971,
SonarSource-sonar-java-295863948, paramiko-paramiko-306104686,paramiko-paramiko-306104687, paramiko-paramiko-306104688, paramiko-paramiko-306104689, paramiko-paramiko-306104690.
We provided this list to \bugswarm's authors, we expect that the missing Docker images will be available in the following weeks.

By only considering the pairs of builds that are available, reproduced  at least five times successfully and based 
on a unique commit we end up with \numprint{1759} pairs of builds taken from 156 GitHub repositories and with 
builds that are \numprint{2.56} years old on average.

\begin{table}[t]
  \centering
  \caption{Number of pairs of builds in \bugswarm.}
  \label{tab:buggy_metric}
  \begin{tabularx}{\linewidth}{@{}X r r r@{}}
    \toprule
    Metrics & Java & Python & All \\\midrule
\# Pairs of builds in \bugswarm's paper & \numprint{1827} & \numprint{1264} & \numprint{3091}  \\
\# Pairs of builds reproduced 5 times & \numprint{1699} & \numprint{1250} & \numprint{2949} (\numprint{95.41}\%)  \\
\# Pairs of builds with unique commit & \numprint{998}  & \numprint{769}  & \numprint{1767} (\numprint{57.17}\%)  \\
\# Docker images not available & \numprint{3} & \numprint{5} & \numprint{8} (\numprint{0.26}\%)  \\
    \bottomrule
  \end{tabularx}
\end{table}

\begin{table}[t]
  \centering
  \caption{Number of pairs of builds that have duplicate content and that change source code. Those metrics are used to verify the requirements one and four for automatic program repair.}
  \label{tab:patch_metric}
  \begin{tabularx}{\linewidth}{@{}X rrr@{}}
    \toprule
    Metrics & Java & Python & All \\\midrule
\# Empty diffs & \numprint{1} & \numprint{1} & \numprint{2} (\numprint{0.11}\%)  \\
\# Duplicate diffs & \numprint{101} & \numprint{97} & \numprint{198} (\numprint{11.21}\%)  \\
\# Duplicate messages & \numprint{178} & \numprint{154} & \numprint{332} (\numprint{18.79}\%)  \\
\# Diffs that change source & \numprint{827} & \numprint{445} & \numprint{1272} (\numprint{71.99}\%)  \\
\# Diffs that only change source & \numprint{305} & \numprint{161} & \numprint{466} (\numprint{26.37}\%)  \\
    \bottomrule
  \end{tabularx}
\end{table}

\begin{table}[t]
  \centering
  \caption{Average and median time for the developers to fix the builds and the execution time of the passing and failing builds. It shows that builds are fixed much more quickly than traditional bugs. The execution time is smaller for failing builds (1min 12 sec) which indicates that the problems in the builds have a significant impact.}
  \label{tab:patch_execution_time}
  \begin{tabularx}{\linewidth}{@{}X rrr@{}}
    \toprule
    Metrics & Java & Python & All \\\midrule
    Avg. fix time & {1 day, 10.5h} & {1 day, 2.8h} & {1 day, 7.1h} \\
    Med. fix time & {28:54} & {39:52} & {32:45} \\
    Avg. execution time passing & {04:43} & {04:54} & {04:48} \\
    Med. execution time passing & {03:55} & {02:57} & {03:27} \\
    Avg. execution time failing & {02:55} & {04:30} & {03:36} \\
    Med. execution time failing & {02:26} & {02:46} & {02:37} \\
    \bottomrule
  \end{tabularx}
\end{table}

\subsubsection{Human Patches}

For the second angle, we look at the human patches. 
We analyze the diffs between the buggy source code and patched source code and the time needed by the human to create a patch. 
For this analysis, we only consider the \numprint{1767} builds that have a unique commit.
Since the builds that have duplicate commits have the same diff, it would produce a bias if we consider them in this analysis.

\autoref{tab:patch_metric} presents the main metrics on the diffs between the buggy and passing builds.
The first row presents the number of empty diffs, i.e.,  no change
between the buggy and the patched source. For one build
\href{https://github.com/web2py/web2py/compare/6f91fdd8332beb5e6a17a1444655e9b9f22e2f4c...587ff56a94fb774609a474ee82c8166a05f31904}{web2py-web2py-61468453},
the diff is empty because the modification consists of a change in the
configuration of a submodule that does not result in a code changes. We did
not find a reasonable explanation for the build \texttt{checkstyle-checkstyle-211109551}. The
original diff in the project repository is not empty, but the diff generated
inside the Docker image is. The second row contains the number of diffs that
are duplicated, i.e., an exact match of the diff according to the md5 hash
function. For the third row, we look at the commit messages of the passing builds and count 
how many of them are unique. The fourth row shows the number of diffs
that change at least one source file, i.e., a Java or Python file. The following
row contains the number of diffs that only change a source file, i.e., it does
not change a configuration file for example.

\autoref{tab:patch_execution_time} presents metrics that are time-related, the first ones present the
average and median time required by the developer to fix their builds. Then, the
average and median execution time of the failing and passing builds.

The takeaways of those tables are that \bugswarm contains duplicate diffs (\numprint{198}) even when we only consider the pairs of builds that have different commit id.
They also use frequently the same commit message (\numprint{332}). For example,
the message ``Added missing javadoc'' occurs 34 times, ``Added hint for
findbugs.'' 27 times, ``Fixed test.'' 19 times, ``Fixed javadoc.'' 17 times,
``Fix build'' 17 times. Most of the pairs of builds modify at least one source
code file but it is less frequent that the developers do not also change a
different type of file.
This indicates that \bugswarm contains similar type of changes that could introduce APR and FL overfitting.
It also shows that \numprint{73.63}\% (\numprint{1301}) of the builds change at least one non-source file. 
Consequently, the techniques have to support multiple file types in order to be evaluated on \bugswarm.

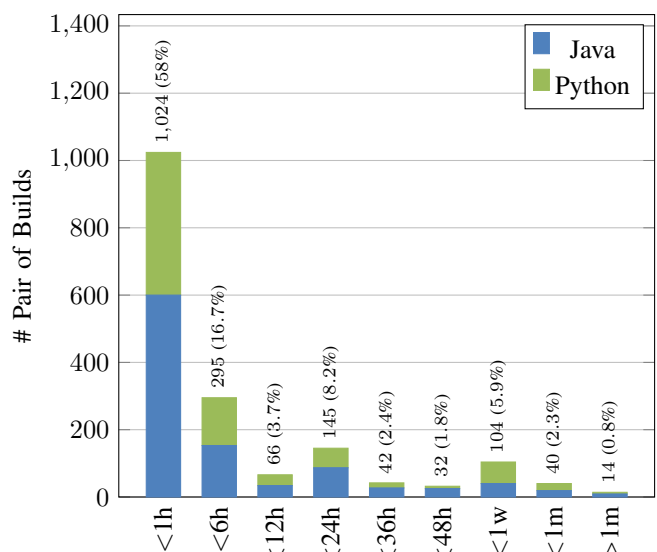
\begin{figure}
  \begin{tikzpicture}
    \begin{axis}[
        show sum on top/.style={
            /pgfplots/scatter/@post marker code/.append code={%
                \node[
                    at={(normalized axis cs:%
                            \pgfkeysvalueof{/data point/x},%
                            \pgfkeysvalueof{/data point/y})%
                    },
                    style={color=black,font=\scriptsize,rotate=90},
                    anchor=west,
                ]
                {\pgfkeys{/pgf/fpu=true}\pgfmathprintnumber{\pgfkeysvalueof{/data point/y}}~(\pgfmathparse{\pgfkeysvalueof{/data point/y}/1767*100}%
    \pgfmathprintnumber[fixed,precision=1]{\pgfmathresult}\%)};
            },
        },
        ybar stacked,
        width  = 0.48*\textwidth,
        height = 8cm,
        major x tick style = transparent,
        bar width=13pt,
        ymajorgrids = true,
        ylabel = {\# Pair of Builds},
        symbolic x coords={$<$1h,$<$6h,$<$12h,$<$24h,$<$36h,$<$48h,$<$1w,$<$1m,$>$1m},
        xtick = data,
        scaled y ticks = false,
        ymin=0,
        nodes near coords={\pgfmathfloatifflags{\pgfplotspointmeta}{1}{}{\pgfmathprintnumber{\pgfplotspointmeta}}},
        every node near coord/.append style={color=black,font=\scriptsize},
        enlarge y limits={value=0.4,upper},
        x tick label style={rotate=90},
        extra y tick labels={},
        extra y tick style={grid=major,major grid style={thick,draw=black}},
    ]
        \addplot[ybar,style={bblue,fill=bblue}]
            coordinates {($<$1h, 600) ($<$6h, 153) ($<$12h, 34) ($<$24h, 87) ($<$36h, 27) ($<$48h, 25) ($<$1w, 40) ($<$1m, 19) ($>$1m, 9) };
        \addplot[ybar,style={ggreen,fill=ggreen},show sum on top]
            coordinates {($<$1h, 424) ($<$6h, 142) ($<$12h, 32) ($<$24h, 58) ($<$36h, 15) ($<$48h, 7) ($<$1w, 64) ($<$1m, 21) ($>$1m, 5) };
        \legend{\strut Java, \strut Python}
    \end{axis}
\end{tikzpicture}
  \vspace{-0.5cm}
  \caption{Amount of time between the buggy commit and the fixing commit for each pair of builds. Legend: `h' means hour, `w' week and `m' month.
  This figure indicates that 86.59\% of the builds have been fixed in less than one day which is much faster compared to an average bug fix time. }
  \label{fig:fix_time}
\end{figure}

The median time to fix a build is low at 33 minutes. Especially, when we
consider the median time that is required to fix a bug like presented by
Valdivia et al. \cite{valdivia2014characterizing}. They indicate that it takes
for eight open-source projects between 35 and 204 days. This highlight, a big
difference between build fixes and bug fixes. For example, we observe that some
bug fixes consist of ignoring or commenting tests to make the build passes. This
type of fix does not fix the regression in the application but keep the build
status in the green. It is the case for example for the build
\href{https://github.com/petergeneric/stdlib/compare/5f6950c71311e3d68f963dc9cdec651bb9ed1353..09730a2d8e40dc36d5fa727c25304f86390ed8eb}{petergeneric-stdlib-160464757}.
\autoref{fig:fix_time} shows the distribution of the fixing time. It indeed
shows that 58\% (\numprint{1024}/\numprint{1767}) of the build are fixed within one hour
and that \numprint{86.59}\% of the builds are fixed in less than one day. Only 54 builds are fixed after one week.

Finally, the failing execution are finishing faster than the passing
build. It takes on average one minute and ten seconds less to execute (\numprint{22.5}\%).
This indicates that the changes between the buggy and passing builds are important.
They impacts significantly the execution time of the failing builds.
Therefore, \bugswarm is a challenging target for APR and FL.

We now focus our analysis on the diff itself. \autoref{tab:count_file_change}
and \autoref{tab:count_line_change} present respectively the total number of
files changed and the total number of changed lines in the \bugswarm benchmark
without considering duplicate commits.

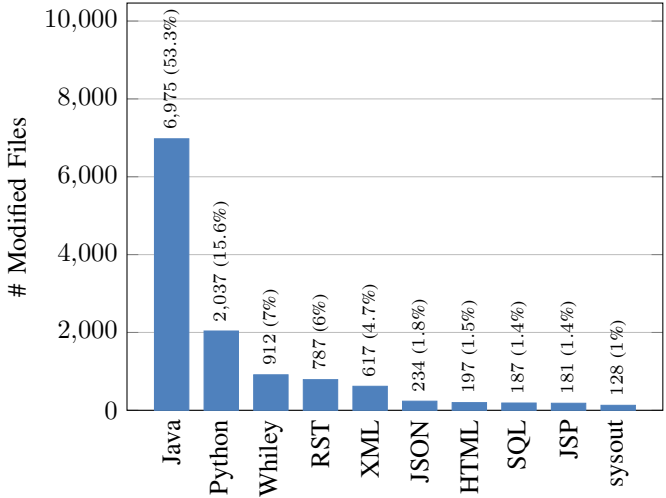
\begin{figure}[t]
  \begin{tikzpicture}
    \begin{axis}[
        show sum on top/.style={
            /pgfplots/scatter/@post marker code/.append code={%
                \node[
                    at={(normalized axis cs:%
                            \pgfkeysvalueof{/data point/x},%
                            \pgfkeysvalueof{/data point/y})%
                    },
                    style={color=black,font=\scriptsize,rotate=90},
                    anchor=west,
                ]
                {\pgfkeys{/pgf/fpu=true}\pgfmathprintnumber{\pgfkeysvalueof{/data point/y}}~(\pgfmathparse{\pgfkeysvalueof{/data point/y}/13077*100}%
    \pgfmathprintnumber[fixed,precision=1]{\pgfmathresult}\%)};
            },
        },
        ybar stacked,
        symbolic x coords={Java,Python,Whiley,RST,XML,JSON,HTML,SQL,JSP,sysout},
        nodes near coords,
        width  = 0.48*\textwidth,
        height = 7cm,
        major x tick style = transparent,
        bar width=13pt,
        ymajorgrids = true,
        ylabel = {\# Modified Files},
        xtick = data,
        scaled y ticks = false,
        ymin=0,
        nodes near coords={\pgfmathfloatifflags{\pgfplotspointmeta}{1}{}{\pgfmathprintnumber{\pgfplotspointmeta}}},
        every node near coord/.append style={color=black,font=\scriptsize},
        enlarge y limits={value=0.5,upper},
        x tick label style={rotate=90},
        extra y tick labels={},
        extra y tick style={grid=major,major grid style={thick,draw=black}},
    ]
        \addplot[ybar,style={bblue,fill=bblue}, show sum on top]
            coordinates {(Java, 6975) (Python, 2037) (Whiley, 912) (RST, 787) (XML, 617) (JSON, 234) (HTML, 197) (SQL, 187) (JSP, 181) (sysout, 128) };

    \end{axis}
\end{tikzpicture}
  \vspace{-0.9cm}
  \caption{Top 10 most frequently modified file types.}
  \label{fig:file_type_change}
\end{figure}

It shows \bugswarm, as expected, that it is mostly existing files that are
modified. Only a small number of files are added and modified.
\autoref{fig:file_type_change} details this analysis by listing the top 10
modified file types. It shows that the main source files of are indeed the most
frequently modified with \numprint{6975} files for Java and \numprint{2037}
files for Python. It highlights the fact that Java projects are more likely to
modify more files than Python projects.

\autoref{tab:count_line_change} shows that the number of added line vs. removed
line in existing files are relatively similar in the total but differ for each language. Indeed, Java diffs contain more added lines than removed
one, and it is the opposite for Python.
It is a good and bad news for automatic program repair and fault localization. It means that the majority 
of the changes are in type of files that are handled by the tools but it also shows that the diffs are big.
Most of the current approaches only handle changes in one location which is not the case in the majority of 
\bugswarm's pairs of builds.

\begin{table}[t]
  \caption{Number of modified, added and removed files in the human patches, considering unique commits. This shows that APR needs multilocation repair ability to target \bugswarm.} \label{tab:count_file_change}
  \centering
  \begin{tabularx}{\linewidth}{@{}X r r r@{}}
    \toprule
    Metric     & Java  & Python & All \\\midrule
\# Modified files & \numprint{9756} & \numprint{3321} & \numprint{13077} \\
\# Added files & \numprint{1086} & \numprint{470} & \numprint{1556} \\
\# Removed files & \numprint{872} & \numprint{49} & \numprint{921} \\
    Avg. \# changed files & \numprint{9.97} & \numprint{4.25} & \numprint{7.71} \\
    \bottomrule
  \end{tabularx}
\end{table}

\begin{table}[t]
  \caption{Number of modified, added and removed lines in the human patches, considering unique commits only.} \label{tab:count_line_change}
  \centering
  \begin{tabularx}{\linewidth}{@{}X r r r@{}}
    \toprule
    Metric     & Java & Python  & All \\\midrule
\# Added lines in modified files & \numprint{431880} & \numprint{182707} & \numprint{614587} \\
\# Removed lines in modified files & \numprint{410086} & \numprint{326831} & \numprint{736917} \\
\# Added lines in added files & \numprint{102097} & \numprint{121607} & \numprint{223704} \\
\# Removed lines in removed files & \numprint{235225} & \numprint{2551} & \numprint{237776} \\
Avg. patch size & 1182 & 824 & 1026 \\
    \bottomrule
  \end{tabularx}
\end{table}

\begin{table*}[t]
  \centering
  \caption{Number of pair of builds for the ten categories of failures considered in this paper.}
  \label{tab:failure_type}
  \begin{tabularx}{\linewidth}{@{}X rrr rrr@{}}
    \toprule
    \multirow{2}{*}{Failure type} & \multicolumn{3}{c}{Duplicate Commit} & \multicolumn{3}{c}{Unique Commit} \\\cline{2-7}
    & Java & Python & Total & Java & Python & All \\\midrule
    Test failure & \numprint{932} (\numprint{54.86}\%) & \numprint{906} (\numprint{72.48}\%) & \numprint{1838} (\numprint{62.33}\%) & \numprint{575} (\numprint{57.62}\%) & \numprint{487} (\numprint{63.33}\%) & \numprint{1062} (\numprint{60.1}\%)  \\
    Checkstyle & \numprint{320} (\numprint{18.83}\%) & \numprint{8} (\numprint{0.64}\%) & \numprint{328} (\numprint{11.12}\%) & \numprint{149} (\numprint{14.93}\%) & \numprint{5} (\numprint{0.65}\%) & \numprint{154} (\numprint{8.72}\%)  \\
    Compilation error & \numprint{263} (\numprint{15.48}\%) & \numprint{33} (\numprint{2.64}\%) & \numprint{296} (\numprint{10.04}\%) & \numprint{167} (\numprint{16.73}\%) & \numprint{16} (\numprint{2.08}\%) & \numprint{183} (\numprint{10.36}\%)  \\
    Doc generation & \numprint{0} & \numprint{171} (\numprint{13.68}\%) & \numprint{171} (\numprint{5.8}\%) & \numprint{0} & \numprint{170} (\numprint{22.11}\%) & \numprint{170} (\numprint{9.62}\%)  \\
    Missing license & \numprint{21} (\numprint{1.24}\%) & \numprint{0} & \numprint{21} (\numprint{0.71}\%) & \numprint{13} (\numprint{1.3}\%) & \numprint{0} & \numprint{13} (\numprint{0.74}\%)  \\
    Dependency error & \numprint{20} (\numprint{1.18}\%) & \numprint{0} & \numprint{20} (\numprint{0.68}\%) & \numprint{12} (\numprint{1.2}\%) & \numprint{0} & \numprint{12} (\numprint{0.68}\%)  \\
    API Regression & \numprint{4} (\numprint{0.24}\%) & \numprint{0} & \numprint{4} (\numprint{0.14}\%) & \numprint{2} (\numprint{0.2}\%) & \numprint{0} & \numprint{2} (\numprint{0.11}\%)  \\
    Unable to clone & \numprint{3} (\numprint{0.18}\%) & \numprint{0} & \numprint{3} (\numprint{0.1}\%) & \numprint{3} (\numprint{0.3}\%) & \numprint{0} & \numprint{3} (\numprint{0.17}\%)  \\
    Missing main file & \numprint{1} (\numprint{0.06}\%) & \numprint{0} & \numprint{1} (\numprint{0.03}\%) & \numprint{1} (\numprint{0.1}\%) & \numprint{0} & \numprint{1} (\numprint{0.06}\%)  \\
\midrule
    Unknown & \numprint{132} (\numprint{7.77}\%) & \numprint{129} (\numprint{10.32}\%) & \numprint{261} (\numprint{8.85}\%) & \numprint{74} (\numprint{7.41}\%) & \numprint{89} (\numprint{11.57}\%) & \numprint{163} (\numprint{9.22}\%)  \\
    \bottomrule
  \end{tabularx}
\end{table*}

Furthermore, we observe that the diffs from the Docker images are not always identical to the original 
diff generated by GitHub between the failing and passing commits.
For example, for the build ansible-ansible-79500861, GitHub sees 494 changes in the file 
\texttt{CHANGELOG.md}\footnote{GitHub diff for ansible-ansible-79500861 builds: \url{https://github.com/ansible/ansible/compare/c747109db9e6bcd7185a3e1e2d451494c035f402..e0a50dbd9287e7fc3d81bfe7fb49972cb4900599}} but none  is visible inside the \bugswarm image.\footnote{Generate the diff inside \bugswarm image \texttt{bugswarm run --image-tag ansible-ansible-79500861  --pipe-stdin <<< "cd; cd build; diff -r failed passed"}}
We did not manage to find an explanation that explains this difference.

\subsubsection{Reasons for Failures}

For the final angle, we analyze the reasons of build failures.
In this study, we analyze the failing execution log to extract the reasons for the failures.
We identify nine different reasons that are presented in \autoref{tab:failure_type}.

\begin{enumerate*}
  \item Test failure, this category contains all the builds that finish with a test failure or a test in error.
  \item Checkstyle, those builds failed because of a checkstyle checker.
  \item Compilation error, the syntax of the code leads to a compilation error or an invalid syntax exception.
  \item Doc generation, the build stops because of an error is detected in the documentation.
  \item Missing license, some files of the build contain invalid or missing license header.
  \item Dependency error, the build did not succeed to download one or several dependencies.
  \item Regression detection, the build introduce regressions in the API compared to the previous version.
  \item Unable to clone, during the build, a submodule did not succeed to be cloned.
  \item Missing main file, the main file to execute the build is missing or is invalid.
  \item Unknown,  the last category contains all the builds that we did not succeed to categorize in one of the ten previous categories.
\end{enumerate*}
Based on this categorization, we observe that test failures are by far the most
common reasons for failure with \numprint{1838} builds that fail due to this
reason. It is followed by checkstyle errors and compilations errors with
respectively 328 and 296 occurrences. The following reason, Doc generation, is only
present in one project: terasolunaorg-guideline. This project contains all the
documentation for the TERASOLUNA Server Framework.

\answer{1}{\textbf{What are the main characteristics of \bugswarm's pairs regarding FL and APR requirements?}
\bugswarm is reported to have \numprint{3091} pairs of builds however 142 of them are not reproduced five times which is in contradiction with \bugswarm's paper.
Consequently, we considered \nbPairs pairs of builds from 156 projects. \numprint{63.3}\% of those builds contain a unique commit id
and are on average \numprint{2.56} years old. \numprint{58}\% of the builds
have been fixed within an hour. The human patches contain on average \numprint{7.71} files changes and a similar amount of addition and removal of lines.
Moreover, \numprint{73.63}\% (\numprint{1301}) of the builds change at least one non-source file.
And the most frequent cause of failure is a failing test case that represents 62\% of the case followed by checkstyle errors and compilation errors.
Considering those numbers, it indicates that APR and FL need to be able to localize and handle multilocation faults and support multiple file types in order to be able to target most of the \bugswarm builds.}

\subsection{RQ2. \bugswarm Execution and Storage Cost}\label{sec:rq2}

In this second research question, we analyze the usage cost of \bugswarm
with a specific focus on execution time and storage.

First of all, we present the workflow of \bugswarm usage.
\begin{enumerate}
\item The first step is to list the available pairs of builds of \bugswarm to
get the Docker Tag ID. This first step requires an access token to \bugswarm's
API.

\item Select the builds to execute, for example, a build from a
specific project that fails due to a checkstyle error.

\item Download and extract the Docker image. This step is handled directly by
Docker.

\item Setup the experiment. This step is project dependent. \bugswarm provides a
folder that is shared with the Docker image (\texttt{\mytilde/bugswarm-sandbox/}
is mapped with \texttt{/bugswarm-sandbox/} inside the Docker image) that is used
copy files and tools from the local machine to the Docker image. This is the
main infrastructure to run an experiment on \bugswarm.

\item Start the Docker image. \bugswarm provides two different execution modes.
The first mode is an interactive one, it creates an ssh connection between the
host and the docker image where one can interact with builds using a command line interface. The second mode
allows providing a command line that will be automatically executed when the
image is started. The second execution mode is more appropriate for a large
scale execution.

\item The final step is the execution of the experiment itself.
\end{enumerate}

\begin{table}[t]
  \caption{Metrics of \bugswarm downloading and storage cost.}
  \label{tab:image_size}
  \centering
  \begin{tabularx}{\linewidth}{@{}X rrr@{}}
    \toprule
    Metrics in GB & Java & Python & All\\\midrule
\bugswarm Docker layer size & \numprint{5107} & \numprint{3813} & \numprint{8921} \\
\bugswarm unique Docker layer size & \numprint{1327} & \numprint{919} & \numprint{2246} \\
Avg. size & 3.01 & 3.05 & 3.03 \\
Download all layers (\numprint[Mbits/s]{80}) & 6d, 7.8h & 4d, 17.3h & 11d, 1.16h \\
Download unique layers (\numprint[Mbits/s]{80}) & 1d, 15.4h & 1d, 3.3h & 2d, 18.8h \\

    \bottomrule
  \end{tabularx}
\end{table}

Based on the described workflow, we now present our analysis of \bugswarm usage
in term of execution time and storage. We identify that step number three is the
step that impacts the most the execution time and the storage required by the
experiment. \autoref{tab:image_size} presents the size of \bugswarm Docker
images. 
The first line shows the total amount of data that has to be downloaded.
According to our observations, the ratio between download size and disk storage is \numprint[x]{2.48} and drops to \numprint[x]{0.41} when considering the duplicate layers.
For example, the image scikit-learn-scikit-learn-83097609 requires to download
\numprint[GB]{3.40} and takes \numprint[GB]{7.06} space on the disk if stored alone but takes \numprint[GB]{1.394} if the shared layers are already downloaded.
Based on this observation, we estimate the total disk space required to \numprint[GB]{3680.45}. 
Note, that this ratio between download size and disk storage has been computed on OSX with Docker 18.09.2. 
The ratio can be different on different os and Docker version.

The second line of \autoref{tab:image_size} presents the total amount of data to download \bugswarm if all the Docker layers are conserved. 
Each Docker image is divided into different layers, the layers
are shared between the different images and consequently reduce the total
amount of data that need to be downloaded. 
Unfortunately, we observe that above \numprint[GB]{350} of Docker images, Docker slows dramatically down the computer and the images have to be removed at a frequent interval to make the computer responsive again. 
However, it increases the total amount of data to download and to decompress, and therefore the execution time of the machine.

The third line presents the average image size, it shows that on average the
Python images are slightly bigger than the Java ones. 
The fourth and fifth lines contain respectively the amount of time to download all the layers and the unique layers with a stable connection of \numprint[Mbits/s]{80}. 
With this connection, it takes between 2 and 11 days to download \bugswarm, depending on the number of Docker layers that need to be downloaded.

Based on those numbers, we estimate the cost to download \bugswarm on an Amazon
Cloud Instance to be \numprint[USD]{45.24}. 
This estimation has been computed by selecting the cheapest virtual machine with \numprint[GB]{16} of RAM.\footnote{Amazon stockade pricing:
\url{https://aws.amazon.com/en/ec2/pricing/on-demand/} visited the 24 April
2019} 
This machine costs \numprint[USD]{0.1664} per hour, the renting of the
machine is \numprint[USD/h]{0.1664} $*$ 2d, 18.8h $=$
\numprint[USD]{11.11}. 

We now estimate the cost of the storage. 
The storage on AWS is \numprint[USD]{0.10} per GB per month.\footnote{Amazon stockade pricing: \url{https://aws.amazon.com/en/ebs/pricing/} visited the 24 April 2019}
There are \numprint[GB]{8921.94} $*$ \numprint{0.41} $=$ \numprint[GB]{3680.45} to store which costs \numprint[USD]{368.05} per month or \numprint[USD]{0.51} per hour. 
Consequently, the storage costs \numprint[USD/h]{0.51} $*$
2d, 18.8h $=$ \numprint[USD]{34.13} to download completely \bugswarm. 
Fortunately, there is no cost to download data from the internet (Docker images).

Those costs do not consider the time required to decompress the images and
the execution of an experiment. 
If we consider a 20 minutes experiment per build, it would result on a total of \numprint[h]{983} of execution which represents an additional
cost of \numprint[USD/h]{0.51} $+$ \numprint[USD/h]{0.1664} $*$
\numprint[h]{983} = \numprint[USD]{666.06}. 
Consequently, we estimate that the starting cost of using \bugswarm on AWS is \numprint[USD]{11.11} $+$
\numprint[USD]{34.13} $+$ \numprint[USD]{666.06} $=$ \numprint[USD]{711.30}.
This cost is the cost of a single execution and does not include the cost of
transferring the data from AWS to a different machine (like transferring to
final results to a local machine).

This costs and the execution time can significantly be reduced by selecting
the pairs of builds to execute. This question
of build selection is discussed in the following research question.

\answer{2}{\textbf{What is the execution and storage cost of \bugswarm?}
The estimated cost to download \bugswarm's pairs of builds on an Amazon Cloud Instance is \numprint[USD]{45.24}, 
plus an estimated additional cost of \numprint[USD]{711.30} to run an experiment of 20 minutes on each build.
This cost is estimated by considering \numprint[GB]{3680.45} of storage required by \bugswarm and the 2 days and 18.80 hours to download it.
This cost is mostly due to the Docker images that improve the reproducibility of the pairs of builds. 
We consider that it is a reasonable overhead if an access to servers like AWS servers is possible but unpractical for consumer grade hardware.}

\subsection{RQ3. \bugswarm for APR and FL}\label{sec:rq3}

In this research question, we are looking at the usage of \bugswarm for the
specific field of automatic program repair and fault localization.

\begin{figure}[t]
  \centering
  \includegraphics[width=0.48\textwidth]{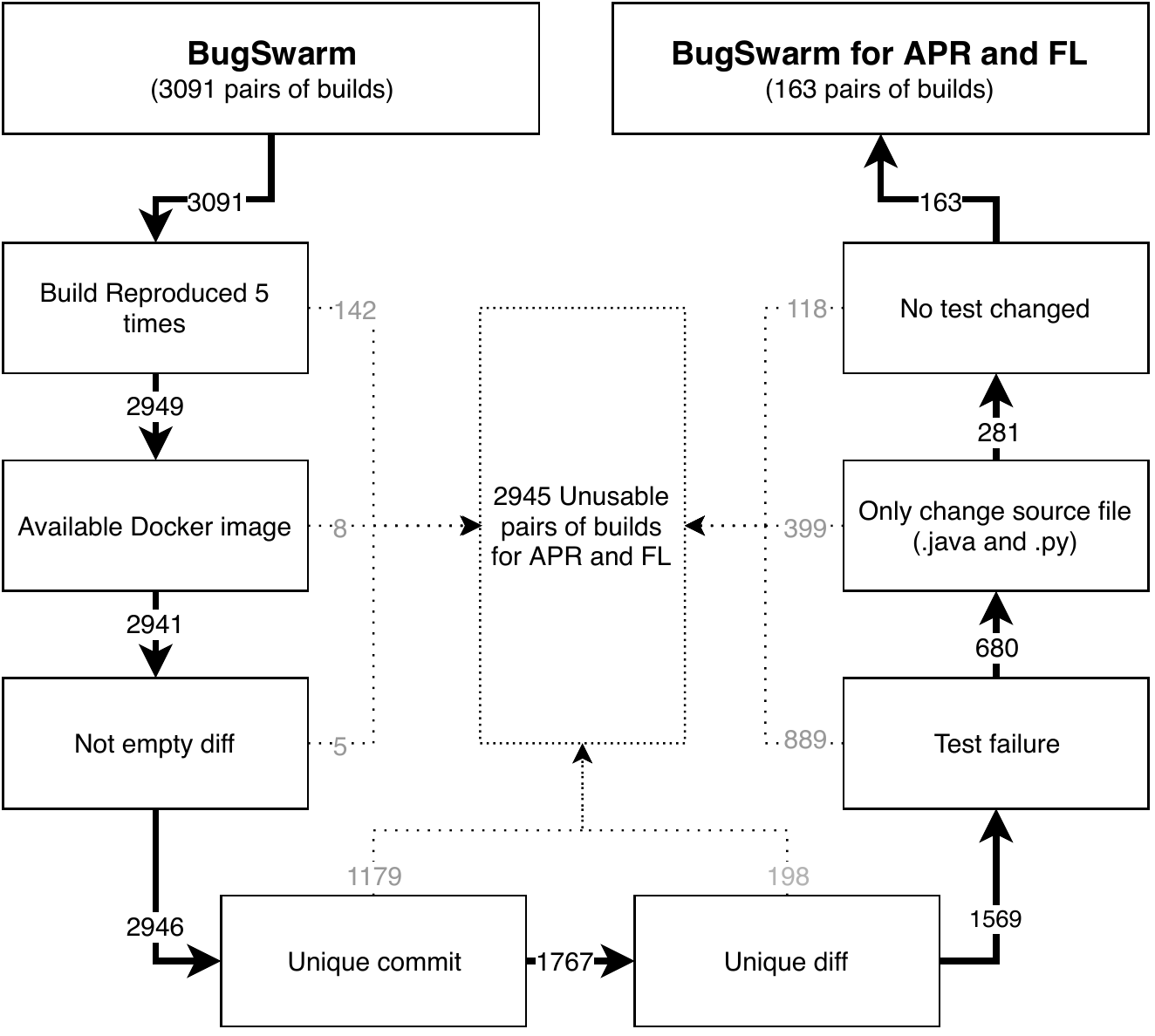}
  \caption{The six filters that select the builds for automatic program repair and fault localization.}
  \label{fig:BugSwarm_filter}
\end{figure}
Based on the requirements presented in \autoref{sec:requirements}, we identify seven filters to select the pairs of
builds that are potentially compatible with automatic program repair and fault
localization. \autoref{fig:BugSwarm_filter} presents the different filters and their impact on the number of pairs of builds.
\begin{enumerate}

  \item \textit{Build reproduced five times.} This filter is presented in \bugswarm's paper~\cite{dmeiri2019bugswarm} to ensure that the pairs of builds are reproducible. 

  \item \textit{Available docker image.} We identify eight Docker images that are missing. We remove them since they cannot be used.
  
  \item \textit{Not Empty diff.} We only consider the pairs of builds are not empty. Since the changes can be related to configuration. The current literature of automatic program repair and fault localization does not target this problem/

  \item \textit{Unique commit.} \numprint{1179} pairs of builds have a
  failing commit that is already present in \bugswarm. We remove those pairs of
  builds since all current approaches are currently working on the source code,
  the same bug will be the same on different pairs of builds which can lead to
  bias in the experiment.

  \item \textit{Unique diff.} This filter is similar to the previous
  one but verifies that the diff of each pair of builds is unique.

  \item \textit{Test-case failure.} The current approaches of automatic program
  repair and fault-localization rely on a failing test-case to expose the bug,
  without a failing test-case those approaches cannot be executed.

  \item \textit{Only change source file.} This filter removes all the pairs of
  builds that modify files that are not source code (.py or .java). We apply
  this filter since to our knowledge no approach is able to handle non-source
  code files.

  \item \textit{No test changed.} The final filter removes the pairs of builds
  that modify a test-case. We remove those builds since the modification of the
  test case change the oracle and therefore the buggy version of the application
  contain either an invalid oracle or not up-to-date one.
\end{enumerate}

After applying the outlined filters on \bugswarm, we are reduced down to 163 pairs of builds, 65 for Java and 98 for Python. Intriguingly, 154 of the 163 pairs of 
builds have been fixed in less than 24 hours (94\% of which have been fixed within one hour).

\begin{figure}[t]
  \begin{tikzpicture}
    \begin{axis}[
        show sum on top/.style={
            /pgfplots/scatter/@post marker code/.append code={%
                \node[
                    at={(normalized axis cs:%
                            \pgfkeysvalueof{/data point/x},%
                            \pgfkeysvalueof{/data point/y})%
                    },
                    style={color=black,font=\scriptsize},
                    anchor=west,
                ]
                {\pgfkeys{/pgf/fpu=true}\pgfmathprintnumber{\pgfkeysvalueof{/data point/x}}~(\pgfmathparse{\pgfkeysvalueof{/data point/x}/163*100}%
    \pgfmathprintnumber[fixed,precision=1]{\pgfmathresult}\%)};
            },
        },
        axis line style={draw=none},
        xbar stacked,
        symbolic y coords={Bug fix,Non-bug fix,Unknown},
        nodes near coords,
        width  = 0.48*\textwidth,
        height = 4.5cm,
        major y tick style = transparent,
        xlabel = {\# Pairs of Builds},
        xtick = data,
        scaled x ticks = false,
        xmin=0,
        enlarge x limits={value=0.3,upper},
        xticklabels={,,,},
        every node near coord/.append style={color=black,font=\scriptsize},
    ]

        \addplot[style={bblue,fill=bblue}]
            coordinates {(3,Unknown) (12,Non-bug fix) (50,Bug fix)};
        \addplot[style={ggreen,fill=ggreen}, show sum on top]
            coordinates {(8,Unknown) (28,Non-bug fix) (62,Bug fix)};

        \legend{\strut Java, \strut Python}
    \end{axis}
\end{tikzpicture}
  \vspace{-.5cm}
  \caption{Type of patches for the 146 pairs of builds compatible with APR and FL.}
  \label{tab:bug_category}
\end{figure}
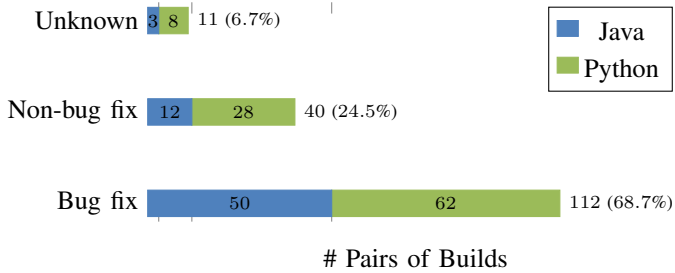

We then manually categorize those 163 pairs of builds into three categories of patches:
\begin{enumerate*}
  \item Bug fix is a patch that we identify as a bug fix
  \item Non-bug fix is a patch that we identify as not a bug fix
  \item Unknown is a patch that we did not succeed to categorize to due to a
  lack of domain knowledge.
\end{enumerate*}
\autoref{tab:bug_category} presents the results of our manual analysis.
It shows that we identify \aprbugs patches that fix a bug, 50 for Java and 62 for Python.
40 pairs of builds have been identified as not a fix for a bug and 11 others as unknown.

We provide the complete list of builds in our repository~\cite{repo} and a
website~\cite{website} that allows browsing, filtering and searching \bugswarm's pairs of
builds. The website illustrates the impact of the
different builds nicely. We recommend the reader to check it to have the
perspective of \bugswarm's content.

\answer{3}{\textbf{Which pairs of builds meet the requirements of
Automatic Program Repair (APR) and Fault Localization (FL) techniques?} We
design eight filters to select the pairs of builds based on the requirements of APR and FL (see \autoref{sec:requirements}). We identify 146 compatible pairs of builds, 81 for Java and 66
for Python which represent \numprint{4.72}\% of the \bugswarm benchmark. On
those 146 pairs of builds, we identify manually 99 bug fix (\numprint{3.2}\%), 48 for Java (\numprint{2.62}\%), 51 for
Python (\numprint{4.03}\%).}

\section{Discussion}\label{sec:discussion}

\subsection{Failing builds vs. bugs}

\bugswarm is a benchmark of pairs of builds, however, its name, \bugswarm, is
misleading. It leads one to think that it is a benchmark of bugs. 
Only by analyzing the content of a benchmark that is possible to realize its true nature.
Indeed, we showed in RQ1 (see \autoref{sec:rq1}) that there are ten different
reasons producing failing builds, such as test failures, checkstyle,
compilation errors. Where only the test failures expose bugs, the other
types of failures do not produce incorrect behavior and therefore should not be
considered as bugs.
For example, the build checkstyle-checkstyle-77722344 fails because of a
checkstyle error: \texttt{Redundant `public' modifier}, as visible in its
execution log: \url{https://travis-ci.org/checkstyle/checkstyle/jobs/77722344},
SonarSource-sonar-java-74910602 fails because it is unable to clone one of its
submodules: \url{https://travis-ci.org/SonarSource/sonar-java/jobs/74910602}.

The confusion between \bugswarm's name and its actual content can lead to
invalid recommendations from reviewers or, much worse, incorrect analysis of
empirical evaluations that use \bugswarm blindly. We would like to bring to
terms the importance of the name especially for artifacts that are designed to
be used for other researchers.

\subsection{Comparison between \bugswarm and existing benchmarks for APR and FL}\label{sec:comparison}

There are a growing number of benchmarks for automatic program repair and
fault localization, for example, Defects4J~\cite{Just2014Defects4J}, Bears~\cite{Madeiral2019Bears},
IntroClass~\cite{LeGoues2015ManyBugsIntroClass},
ManyBugs~\cite{LeGoues2015ManyBugsIntroClass},
IntroClassJava~\cite{Durieux2016IntroClassJava},
Bugs.jar~\cite{Saha2018BugsDotjar}, QuixBugs~\cite{Lin2017QuixBugs},
BUGSJS~\cite{gyimesi2019bugsjs}. \bugswarm shares with
Bears~\cite{Madeiral2019Bears} the \textit{modus operandi} used to collect the
data in the benchmark, they both use TravisCI builds as source.

\bugswarm and Bears are the most similar benchmarks, both of them use TravisCI
to collect builds. Bears~\cite{Madeiral2019Bears} focuses on bugs by reproducing
the build and manually analyzed the human patch, whereas \bugswarm focuses on
reproducible pairs of builds in a more generic way. \bugswarm's infrastructure
is unique: it is the only benchmark that encapsulated each artifact in a Docker
image. This infrastructure comes with the advantage of an improved
reproducibility. However, it will still suffer from unreproducible builds due to
invalid dependencies, for example, a snapshot version can be updated with
breaking changes that will break the compilation of a \bugswarm artifact.
Indeed, the Docker image does not include the dependencies of the application
which can lead to missing or invalid dependencies. This problem is the most
frequent source of unreproducibility of builds according to \bugswarm's
authors. Docker images come also with disadvantages such as a considerable size
and execution overhead, as shown in RQ2 (see \autoref{sec:rq2}). For the
readers' reference, the repository~\cite{repo} that contains all the sources of
all the builds is less than \numprint{2}GB, compared to \numprint{3680.45}GB of
\bugswarm.

The \bugswarm diffs between the failing and passing version are also different
from the existing benchmark. Sobreira et al. \cite{sobreira2018dissection}
observe that the average patch size in Defects4J is 4 lines. Madeiral et al.
\cite{Madeiral2019Bears} report a patch size of 8 lines in Bears. In \bugswarm,
we observe an average patch size of \numprint{1026} lines. This difference of size
highlights the difference of nature of \bugswarm compared to the other
benchmarks.

To sum up, the difference between \bugswarm and the literature is threefold:
\begin{itemize}
  \item A benchmark of reproducible builds that can support new researches on
  build repair;
  \item The diffs are much bigger than the literature;
  \item A novel infrastructure to store and interact with bugs.
\end{itemize}

\subsection{Lessons learned}\label{sec:lessons_learned}

\newcommand*\rot{\rotatebox{90}}
\newcommand*\ok{\ding{51}}
\newcommand*\half{\textbf{\mytilde}}

\begin{table}
  \caption{Recommendations vs. Benchmarks (\ok: satisfied; \half: partially satisfied).}\label{tab:lesson_perspective}
  \begin{tabularx}{\linewidth}{@{}X cccccccccc@{}}
    \toprule
    Lessons learned & 
    \rot{Defects4J} & %
    \rot{Bugs.jar} &  %
    \rot{Bears} &      %
    \rot{QuixBugs} &  %
    \rot{iBugs} &     %
    \rot{ManyBugs} &  %
    \rot{IntroClass} & %
    \rot{IntroClassJava} & %
    \rot{BUGSJS} &    %
    \rot{\textbf{\bugswarm}} \\% 10 
    \midrule
Target       & \ok & \half & \ok & \half & \half & \half & \half & \half & \half & \half \\
Exploration  & \ok &       & \ok &       &       &       &       &       &       &       \\
Dependencies & \ok &       &     &       &       &       &       &       &       &       \\
Access       & \ok & \half & \ok & \half & \half & \half & \half & \half & \half & \half \\
Documentation& \ok &       & \ok &   \ok &       &       &       &       &       & \half \\
Versioning   &     &       & \ok &       &       &       &       &       &       &       \\
    \bottomrule
  \end{tabularx}
\end{table}

Our analysis of \bugswarm allowed us to understand what constitutes a reasonable
benchmark that is suited to fault localization and automatic program repair. In
this section, we discuss several recommendations on how to build and make
available benchmarks of bugs. Also, we draw some conclusions on how to improve
\bugswarm.

\begin{enumerate}
  \item \textbf{[Target]} The first recommendation is to think of the
  requirements of the research fields that the benchmark target. In this case,
  APR and FL require bugs to be exposed by a failing test case as well as
  metadata about each bug such as the location of the source and/or the
  binaries. A benchmark that fails to provide such information is ill-suited to
  fault localization  and automatic program repair;

  \item \textbf{[Exploration]} The second recommendation is to provide the diff
  of each artifact. The diff can be easily understood by humans and give a quick
  understanding of the content of the artifact. We also recommend providing a
  selection of artifacts for each research field if all the artifacts are not
  compatible with the research field. This ensures that the same selection is
  used across papers and guaranty a fair comparison between approaches;

  \item \textbf{[Dependencies]} The next recommendation is to improve the
  reproducibility of the benchmark by including the dependencies of your
  artifacts since it is the main reason for unreproducibility (as pointed out
  by the authors of \bugswarm);

  \item \textbf{[Access]} Provide full access to your benchmark and metadata.
  Benchmarks are by nature an artifact that should be used in different research
  works to compare against other related solutions. Furthermore, Offer the tools
  used to create the benchmarks as open source -- it is required to ensure the
  representativeness of the benchmarks;

  \item \textbf{[Documentation]} Make sure to have proper documentation for your
  benchmark that explains how to checkout one single bug but also how to execute
  a large scale experiment on it. And avoid ``coming soon'' messages or at least
  provide an email address with that message.;

  \item \textbf{[Versioning]} When benchmarks evolve, it is important to version it, to be able to always point out a previous version that was used in a specific experiment.
  Indeed, without a version number it is difficult for authors to refer to which version of the benchmark and therefore readers cannot know which artifacts have been used.
  We recommend that each time a new artifact is added to the benchmark, a new version is created like it is done in Bears benchmark~\cite{Madeiral2019Bears}.
\end{enumerate}

\autoref{tab:lesson_perspective} puts into perspective the lessons learned from this section with nine other existing benchmarks of bugs. 
We observe that none benchmark meets all our recommendations.
However, Defects4J~\cite{Just2014Defects4J} and Bears~\cite{Madeiral2019Bears} already meet 5/6 of our recommendations.

\subsection{Threats to Validity}
As any implementation, the scripts that we use to collect \bugswarm's builds and
the metrics are potentially not free of bugs. A bug might impact the results we
reported in \autoref{sec:contributions}. However, the script and the raw output
are open-source and publicly available for other researchers and potential users
to check the validity of the results.

Moreover, \bugswarm could also be impacted by potential bugs and the benchmark
itself can be updated in the future. Therefore, the observed result can differ
in the future. However, we provide all the scripts that are required to redo
this analysis and update if needed. Moreover, the website and the repository can
be updated to take into account the changes in \bugswarm.

This analysis focuses on the nowadays requirements of automatic program repair
and fault localization techniques. Those requirements can evolve with the time,
and therefore the bug selection presented in RQ3 (see \autoref{sec:rq3}) can be
inadequate in the future.

\subsection{Discussion with \bugswarm's Authors}

We communicated our results with \bugswarm's authors to get their feedback and opinion on it. 
It engendered a really interesting discussion about their vision and future directions for the 
benchmark.
On the one hand, their feedback gave us the opportunity to improve and clarify the paper. On 
the other hand, our work allowed them to identify and fix issues in \bugswarm such as: duplicate 
artifact id, inconsistent number of pairs of builds, unavailable Docker images.
We would like to thank them for their responsiveness and feedbacks.

\section{Related Works}\label{sec:rw}

\subsection{Benchmarks}

We first present the benchmarks for automatic program repair and fault
localization from the literature. The literature contains several benchmarks of
Java bugs. Defects4J~\cite{Just2014Defects4J} is the most used benchmark for
automatic program repair and fault localization. It contains 395 minimized bugs
from six widely used open source Java projects. It has been created by mining
Apache issue tracker. Bugs.jar~\cite{Saha2018BugsDotjar} contains
\numprint{1158} bugs from eight Apache projects. It was created using the same
strategy than Defects4J. IntroClassJava~\cite{Durieux2016IntroClassJava}
contains \numprint{297} bugs from six different student projects. It is a
transpiled version to Java of the bugs from the C benchmark
IntroClass~\cite{LeGoues2015ManyBugsIntroClass}. Bears~\cite{Madeiral2019Bears}
contains 251 bugs from 72 different GitHub projects. It was created by mining
TravisCI builds. And iBugs~\cite{Dallmeier2007} which contains 390 Java bugs.

The literature also contains benchmark of C programs such as:
SIR~\cite{do2005supporting} is a benchmark of seeded faults from nine small to
medium-scale programs in C language.
ManyBugs~\cite{LeGoues2015ManyBugsIntroClass} contains \numprint{185} bugs from
nine open-source C programs. IntroClass~\cite{LeGoues2015ManyBugsIntroClass}
contains \numprint{572} bugs from six student programs.

We decided to analyze \bugswarm instead of the other benchmark of the literature
since \bugswarm is a new benchmark that will enjoy good visibility through ICSE
conference and also because its creation process and its content are different
from other benchmarks.

\subsection{Benchmark Analysis}

The literature contains some studies on existing benchmarks of bugs.
defect characteristics: defect importance, complexity, independence, test
effectiveness, and characteristics of the human-written patch.
Sobreira et al. \cite{sobreira2018dissection} also analyze Defects4J but focus on the
identification of repair actions and repair patterns. Madeiral et al.
\cite{madeiral2018towards} automatize the extraction of repair actions and
repair patterns from diff.
Wang et al. \cite{wang2019attention} present a study that analyzes the impact of ignoring
the project Mockito from empirical evaluations that use Defects4J. They show that
automatic program repair techniques have poorer performance on Mockito and
ignoring it can introduce misleading results. It highlights the importance of
the selection of the artifacts from benchmarks.

Some benchmarks also include an analysis of their content.
iBugs~\cite{Dallmeier2007}, Codeflaws~\cite{Tan2017} and
Bears~\cite{Madeiral2019Bears} contain annotated bugs  with size and syntactic
properties on their patches. ManyBugs~\cite{LeGoues2015ManyBugsIntroClass}
analyses the patches and annotate the bugs when functions, loops, conditional
and function calls were added or when function signatures are changed. Those
analyses are comparable to the analysis included in \bugswarm paper even if the
results of the analysis a different like presented in \autoref{sec:comparison}.

This paper presents an analysis using different metrics such as fix time, size
of the benchmark, execution time, cost, or failure type. Moreover, it also
includes an analysis of the content of the benchmark regarding the research field
that it targets.

\section{Conclusion}\label{sec:conclusion}

This paper presents an analysis of the \bugswarm benchmark. We start off by
analyzing the pairs of builds, the human patches, and failure reasons. We observed
that 142 pairs of builds do not match \bugswarm's including criteria (has to be
reproduced five times), \numprint{1182} pairs of builds contain duplicate commits, and 8
pairs of builds are no longer available.  The human patches modify on average
\numprint{1026} lines in \numprint{7.71} files. The failing builds fails for
\numprint{62.32}\% due to a test, \numprint{11.12}\%  due to a checkstyle error
and \numprint{10.03}\% for a compilation error.

We then analyzed the overhead introduced by the \bugswarm infrastructure compared to a traditional repository. 
We estimate the overhead as \numprint[USD]{45.24}, 2d, 18.8h and \numprint[GB]{3680.45}. 
This is the costs of an improved reproductivity.

Finally, we analyzed \bugswarm with the optic to use \bugswarm for automatic
program repair and fault localization. We identify six requirements that are
needed in order to be able to use the pairs of builds in such domains:
availability, uniqueness of the commit, uniqueness of the diff, source code
based, test failure, test-suite not modified. We manually identified 163
potentially relevant pairs of builds, and we determined that \aprbugs of them
are bug fixes, 40 non-bug fixes and 11 unknowns.

\bugswarm has been presented as a benchmark of pairs of builds for automatically
program repair and fault localization but only \aprbugs/\numprint{3091} (\numprint{3.6}\%)
are compatible with the current automatic program repair, and fault localization
approaches --- a number that falls short of other related benchmarks. Moreover,
\bugswarm's name is confusing, it makes one think that the
benchmark contains bugs, but it is a benchmark of builds which is
conceptually rather different.

During our study, we have collected a number of findings on the requirements that
make a benchmark well suited for fault localization and program repair. We have
discussed this in detail, paving the way for upcoming benchmark sets.

\section*{Acknowledgments}

This material is based upon work supported by Funda\c{c}\~ao para a Ci\^encia e
a Tecnologia (FCT), with the reference PTDC/CCI-COM/29300/2017.

\balance
\bibliographystyle{IEEEtran}
\bibliography{references}

\end{document}